# Super-resolution enabled widefield quantum diamond microscopy

*Feng Xu[1], Jialong Chen[2,4], Yong Hou[1], Juan Cheng[3], Tony KC Hui[3], Shih-Chi Chen[2,4]\*, Zhiqin Chu[1,5]\**

[1]Department of Electrical and Electronic Engineering, The University of Hong Kong, Pokfulam Road, Hong Kong, China

[2]Department of Mechanical and Automation Engineering, The Chinese University of Hong Kong, Shatin, Hong Kong

[3]Master Dynamic Ltd., Hong Kong, China

[4]Hong Kong Centre for Cerebro-Cardiovascular Health Engineering, Hong Kong Science Park, Shatin, Hong Kong.

[5]Joint Appointment with School of Biomedical Sciences, The University of Hong Kong, Pokfulam Road, Hong Kong, China

\*Corresponding authors: S. C. C. (scchen@mae.cuhk.edu.hk) and Z. Q. C. (zqchu@eee.hku.hk).

KEY WORDS: widefield quantum diamond microscopy, super-resolution, digital micromirror device, structured light illumination, fluorescent nanodiamonds

**Abstract:** Widefield quantum diamond microscopy (WQDM) based on Kohler-illumination has been widely adopted in the field of quantum sensing, however, practical applications are still limited by issues such as unavoidable photodamage and unsatisfied spatial-resolution. Here, we design and develop a super-resolution enabled WQDM using a digital micromirror device (DMD)-based structured illumination microscopy. With the rapidly programmable illumination patterns, we have firstly demonstrated how to mitigate phototoxicity when imaging nanodiamonds in cell samples. As a showcase, we have performed the super-resolved quantum sensing measurements of two individual nanodiamonds not even distinguishable with conventional WQDM. The DMD-powered WQDM presents not only excellent compatibility with quantum sensing solutions, but also strong advantages in high imaging speed, high resolution, low phototoxicity, and enhanced signal-to-background ratio, making it a competent tool to for applications in demanding fields such as biomedical science.

Introduction

Diamond-based quantum sensing has emerged as a promising technology, which utilizes ultrastable point defects to measure physical quantities with unprecedented sensitivity and precision at nanoscale. [1-4] The nitrogen-vacancy (NV) center hosted in diamond is one of the most studied quantum sensors due to its ultra-long coherence time and easy spin manipulation under ambient conditions. [2, 5, 6] Moreover, the superior properties of diamond materials, such as chemical inertness, flexible modality, and excellent biocompatibility, have promoted the development of NV-based metrology in various scientific fields, including life science, material science, geoscience, physics and engineering. [2, 7, 8] General optical platforms for quantum sensing include confocal and wide-field fluorescence microscopy. Compared with a confocal system [8], a wide-field system can image and monitor an ensemble of NV centers within a large field of view in parallel.[9-11] Conventional wide-field fluorescence microscopy normally adopts the Kohler illumination to obtain a fixed light pattern over the entire sample plane.[12] Nevertheless, long-term laser exposure might cause severe problems for many photosensitive samples such as living cells,[13,14] making live-cell imaging difficult due to the unavoidable phototoxicity. To enhance imaging resolution and reduce phototoxicity, the concept of programmable illumination via adaptive optics has been developed.[15,16] For example, real-time modulation of illumination patterns has been applied to improve imaging speed, quality[17,18] and sensing sensitivity,[16] especially for biological samples.[19] Artificial intelligence (AI) technology, which has been developed rapidly in recent years, has also been successfully applied to adaptive optics, significantly improving the imaging results.[16]

Another notable point is that the diffraction-limited spot size[20] in conventional optical systems often prevents an imaging system from resolving information from tightly packed emitters. This hinders the exploration of WQDM for the advanced study of emerging materials such as magnetic skyrmions,[21] and intracellular temperature sensing[22] where high spatial resolution is required. Recently, several works have attempted to realize super-resolution in WQDM, for example, through stochastic optical reconstruction microscopy (STORM);[23,24] and great efforts have also been made to integrate super-resolution imaging with quantum sensing measurements.[25,26] For example, by modulating the fluorescence intensity of the NV centers within nanodiamonds via electron spin resonance (ESR), a spatial resolution of 12 nm was achieved in a quantum sensing setup in 2013.[25] Nevertheless, this approach cannot be extended to general quantum sensing platforms due to the coupling of imaging technique to NV-based quantum measurements. On the other hand, structured illumination microscopy (SIM), which can achieve sub-diffraction limit imaging via a wide-field system, has been widely used in biology owing to its simplicity in operation and low photo-damage to samples. [24, 27-29] In general, a two-fold spatial resolution improvement can be achieved by shifting the unobservable high spatial frequency information to an observable low spatial frequency region via structured light illumination (SLI). [30, 31] Commonly used SLI devices include optical gratings, [32] liquid crystal-based spatial light modulators (SLMs) [33] and digital micromirror devices (DMDs). [29] Among these devices, the DMD has the advantages of high speed (up to 32.5 kHz), high resolution (e.g., 3840 × 2160 pixels), and low cost, and has therefore been frequently used in high-speed imaging systems. [34, 35] For example, by combing a DMD with digital holography, our team has demonstrated super-resolution imaging (155 nm) at an imaging rate of 26.7 frames per second. [29]

Built upon our experience in DMD-based light microscopy, [34, 36-45] by integration with conventional diamond quantum sensing, we present a dynamic super-resolved WQDM. Firstly, we experimentally show SIM-based super-resolution imaging of fluorescent nanodiamonds (FNDs) with a lateral sub-diffraction resolution of 202 nm. Importantly, in the experiment we found no interference by implementing the DMD-based SIM in our quantum sensing schemes. Finally, as a proof-of-concept demonstration, we obtained the super-resolved ODMR measurements of two FNDs physically separated by 83 nm, which was not possible via conventional WQDM.

Results

To construct the dynamic WQDM, a DMD was included in the excitation optical path in the system to generate programmable illumination patterns, as shown in Figure 1a. Based on the system, the ensemble NV centers can be modulated using microwave, facilitating the widefield quantum sensing measurements. It is worth noting that the implementation of our DMD-based device is simple and compatible with most of the existing WQDM apparatus. The DMD can generate arbitrarily designed light fields on the surface of the diamond sample, i.e., a programmed illumination area of the HKU logo, as shown in Figure 1a.

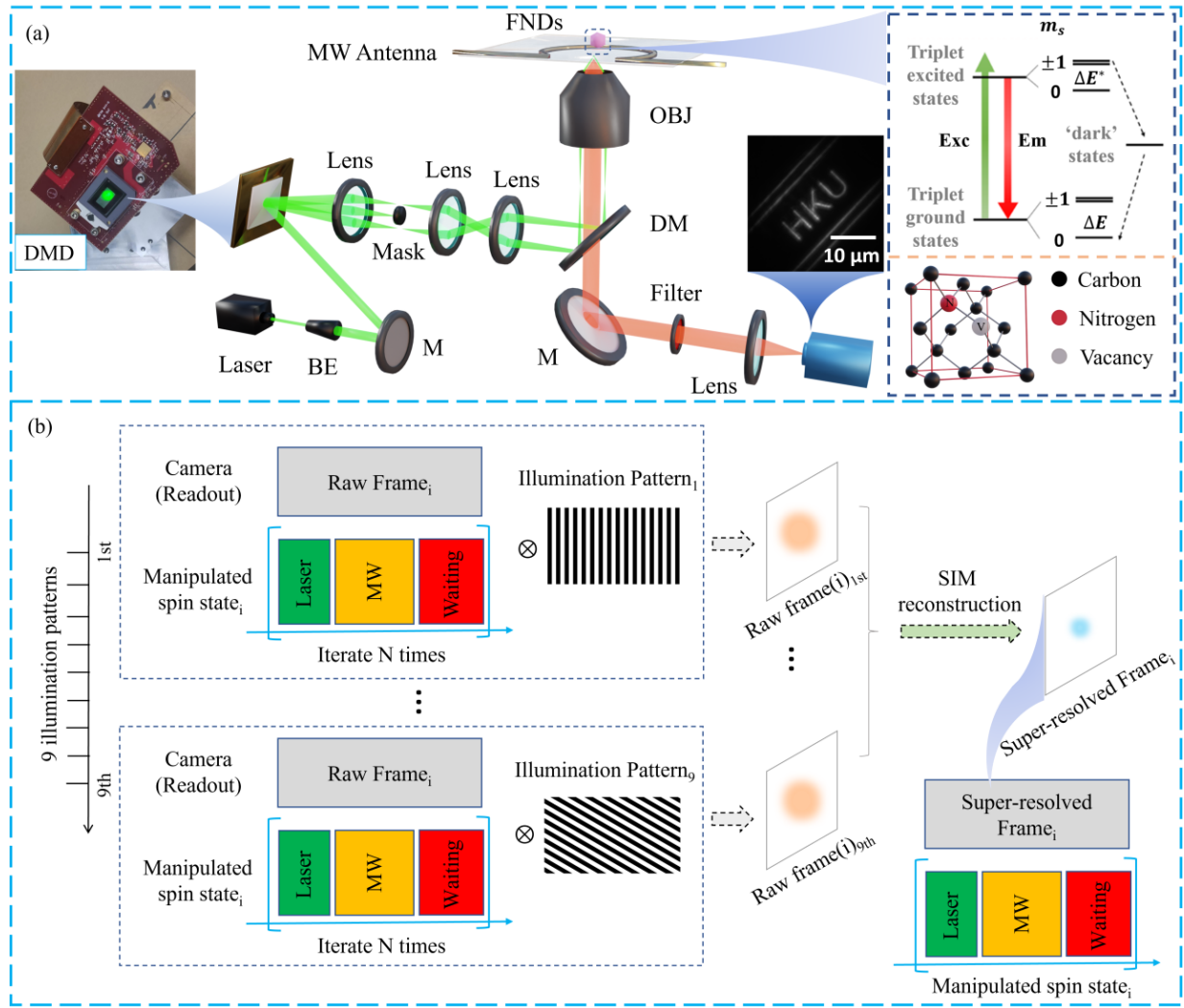

**Figure 1.** Design and working principle of dynamic widefield quantum diamond microscopy. (a) Optical configuration of the measurement setup. DMD: digital mirror device; DM: dichroic mirror; OBJ: objective. The insert shows the crystal structure of a single NV center and the energy structure. A customized illumination pattern of the HKU logo generated by the DMD. (b) Measurement protocols to achieve the general super-resolved quantum sensing scheme based on the DMD.

We have developed a super-resolution imaging function using DMD-based SIM, which can be seamlessly integrated with conventional WQDM for super-resolved widefield quantum sensing. This excellent compatibility with previously developed quantum sensing schemes is one of the most attractive features of the current method (Figure 1b). For any of the spin states, we can record nine raw frames under successive illumination patterns and reconstruct one super-resolved frame based on the algorithm for DMD-based SIM. [31] We can also arbitrarily tune the parameters (e.g., duration of laser, MW and/or waiting time) in the quantum measurement sequence to implement any of the desired protocols, such as Rabi, $T_1$, etc.

Next, we demonstrate how programmable illuminations enabled by the DMD can improve the performance (Figure 2a) of monitoring FNDs in 3T3 cells samples using our customized WQDM. In conventional illumination with a Gaussian beam, we have noticed that some FNDs were extremely bright (approaching the upper detection limit of the camera), possibly due to their well-known aggregation in cytosol. [46] As a result, this extremely high photon count for a very tiny portion (only 0.68% of the total illumination area) of the region of interest (ROI), will in fact limit the allowed laser power over the whole field of view. If one was interested in studying the FNDs located within the cell periphery (as indicated by white arrows), investigation of this ROI would be practically impossible owing to insufficient excitation laser power (Figure 2b). With the introduced dynamic feature of WQDM, however, we could now "turn off" the unwanted bright spots to increase the laser power to focus on our targeted ROI (Figure 2). This has enabled us to acquire more photon counts from the targeted ROI by increasing the excitation laser power. (Figure 2c). As shown in the intensity profile in Figure 2e, the fluorescence intensity of the covered FNDs decreased from about 65000 to 20000, while the intensity of the dark FND increased from 7000 to 17000 after the laser power increased. Next, for mitigating the unwanted effects of laser on the cells, we further showed a customized illumination pattern to "turn off" the direct illumination on the cell body (Figure 2d&e).

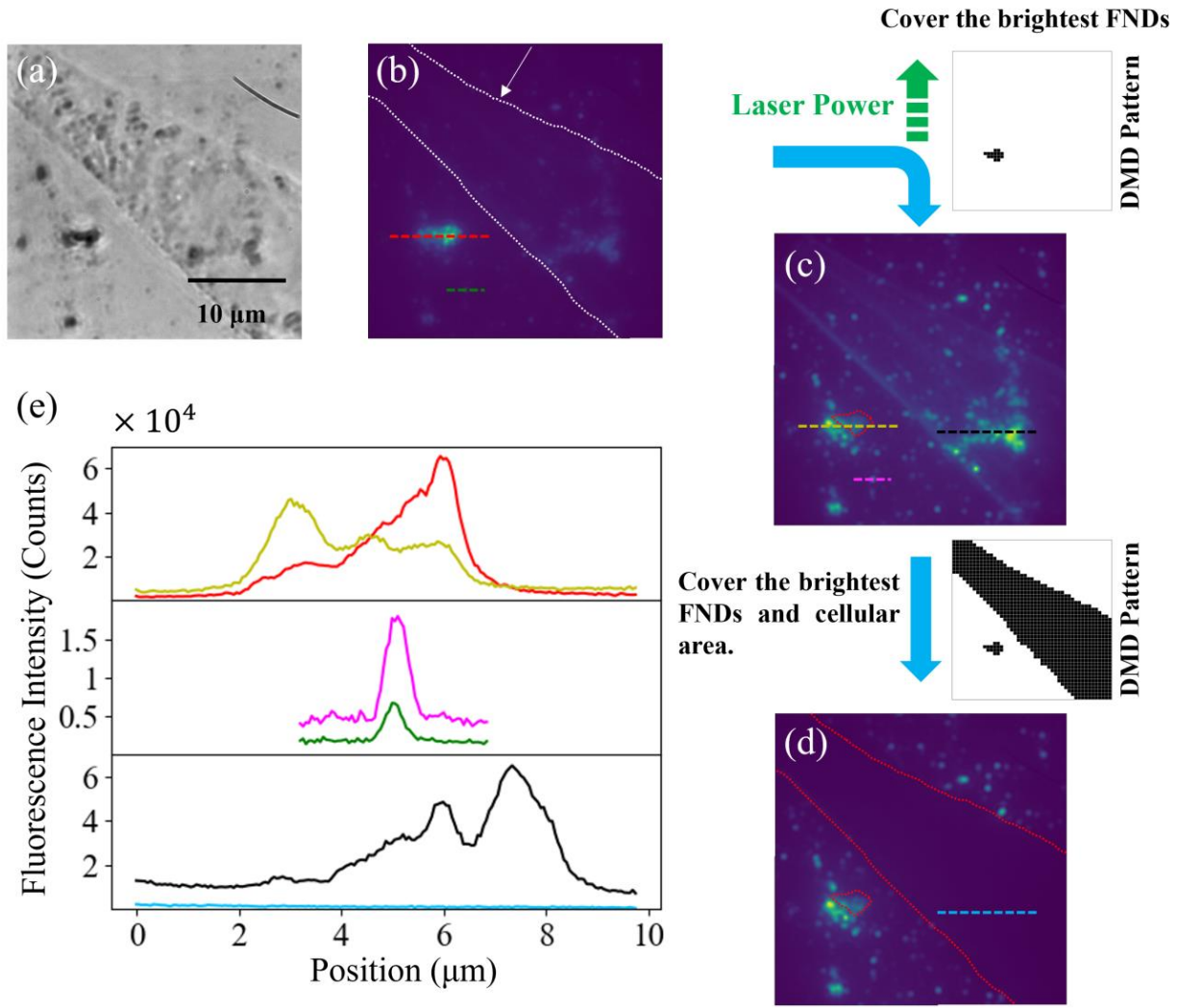

**Figure 2.** Demonstration of a versatile illumination concept in dynamic WQDM. (a) Transmitted light image of a cell fed with FNDs. Fluorescence images of the cell via (b) conventional illumination, (c) and (d) customized illumination. (e) Corresponding line profiles of the dashed lines. The corresponding customized illumination patterns have been shown in the DMD patterns, where the black area indicates the turned-off area on the DMD.

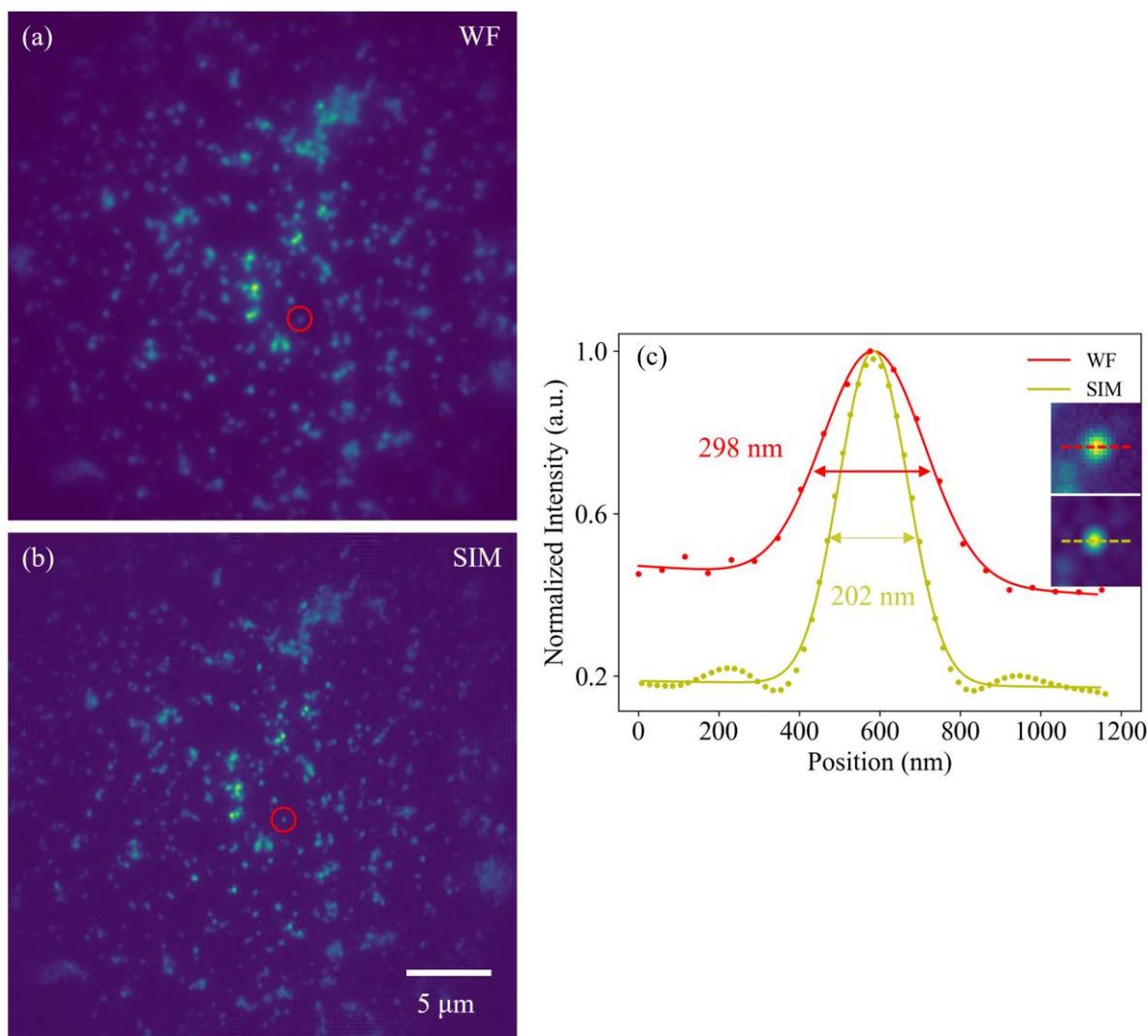

**Figure 3.** SIM-enabled super-resolution imaging of 30 nm FNDs. (a) and (b) Conventional WF image and SIM image. (c) Line profile of the fluorescence intensity from conventional WDQM and super-resolved WDQM. The fluorescence line profiles are fitted with Gaussian curves.

Figure 3 compares the spatial resolution of the conventional wide-field microscope and the structured illumination microscope. The original wide-field image and reconstructed super-resolved image based on SIM are shown in Figure 3a and b, respectively. An FND with nominal diameter of 30 nm has been selected as a point light source for calibrating the spatial resolution of our customized microscope. According to the selected FNDs, labelled by dashed lines on the insert of Figure 3c, the resolution of home-built wide-field microscope is 298 nm (red line in Figure 3c).

The theoretical spatial resolution of wide-field microscope is 241 nm, based on the Abby criterion, with the emission at 700 nm. After the SIM reconstruction, the spatial resolution has been improved to 202 nm, breaking the diffraction limit as shown in the yellow line in Figure 3c (the theoretical spatial resolution based on SIM is 159 nm [31]). For the FNDs sample we used here, the signal-background ratio (SBR) has been improved from 2 (WF) to 5 (SIM) after the super-resolved reconstruction. This effectively enhanced the performance of quantum sensing to a comparable extent.

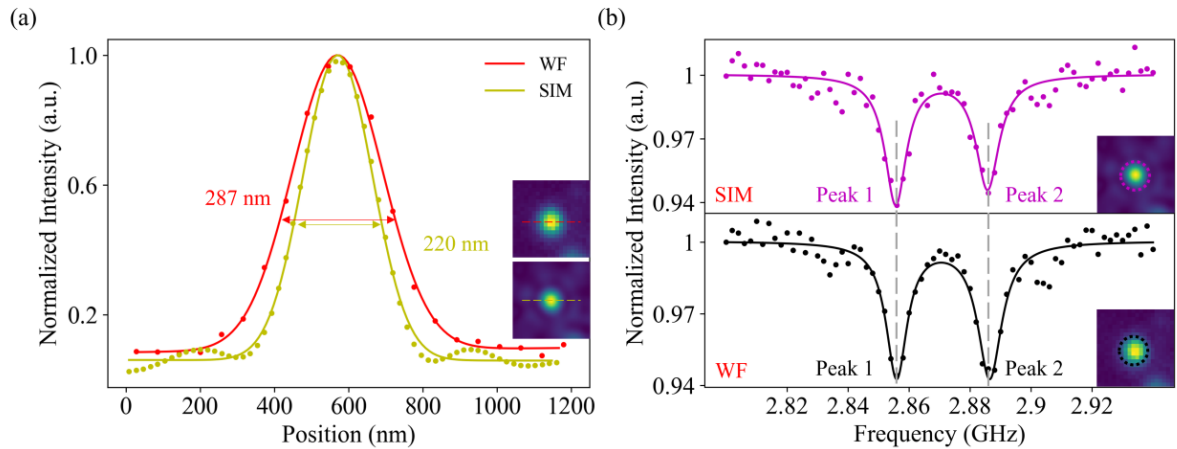

**Figure 4.** Investigating the influence of introduced SIM on the ODMR measurements. (a) Line profiles of the fluorescence intensity from conventional WDQM and super-resolved WDQM. The fluorescence line profiles are fitted with Gaussian curves. (b) The ODMR spectrums extracted from the same FND under the conventional WDQM (lower panel) and super-resolved WDQM (upper panel).

To investigate the influence of the SIM reconstruction process on the ODMR measurement, a single FND with a FWHM of the fluorescence intensity profile at 287 nm under the conventional WDQM was selected, while its FWHM under the super-resolved WDQM has been improved to 220 nm (Figure 4a). The ODMR spectrums in conventional WDQM and super-resolved WDQM have the same resonance frequencies, $2856 \pm 1$ MHz and $2886 \pm 1$ MHz, respectively. The contrast and FWHM of the two resonance peaks in conventional WDQM are 5.7 % and $8.4 \pm 1.8$ MHz, and 5.7% and $9.6 \pm 2.0$ MHz, and for super-resolved WDQM are 6.1% and $8.6 \pm 1.9$ MHz, and 5.4% and $9.3 \pm 2.3$ MHz, respectively. Given the 95% confidence interval of the two-peak Lorentz fitting, these results indicate that the SIM reconstruction process does not interfere with the ODMR measurement.

**Table 1. The FWHM and contrast of each ODMR peaks in Figure 4**

|  |  | Conventional WDQM | Super-resolved WDQM |
|---|---|---|---|
| Peak 1 | Resonance frequency (MHz) | $2856 \pm 1$ | $2856 \pm 1$ |
|  | FWHM (MHz) | $8.4 \pm 1.8$ | $8.6 \pm 1.9$ |
|  | Contrast (%) | 5.7 | 6.1 |
| Peak 2 | Resonance frequency (MHz) | $2886 \pm 1$ | $2886 \pm 1$ |
|  | FWHM (MHz) | $9.6 \pm 2.0$ | $9.3 \pm 2.3$ |
|  | Contrast (%) | 5.7 | 5.4 |

Notes: The fitting results are represented with 95% confidence interval of the Lorentz fit.

Two FNDs (Figure 5a) within the diffraction limit were selected for demonstrating the super-resolved ODMR measurement. Although the centroid positions of these two particles are 330nm apart, the fluorescence intensity peaks were only separated by 238 nm, which was larger than the lateral resolution of SIM and smaller than the lateral resolution of a normal wide-field microscope, as shown in the line profile in Figure 5b. This SIM based super-resolved ODMR method, which we have christened SIM-ODMR, can distinguish the contribution for ODMR from different FNDs, which was not possible for a normal wide-field microscope. In Figure 5c, the ODMR spectrums with external magnetic fields are extracted under the SIM based quantum sensing microscope and conventional quantum sensing microscope. There are six different ODMR spectrum peaks under the wide-field microscope, and it is not possible to assign the ODMR spectrum peaks to their corresponding FND using such a microscope, due to the limit lateral resolution. Using the SIM-ODMR method, by contrast, the ODMR spectrums can be easily assigned to their corresponding FND, as shown in Figure 5c. For each ODMR peak under the SIM-ODMR method, its corresponding ODMR peak using a normal wide-field microscope can be found, demonstrating that the SIM reconstruction process has no influence on the resonance frequency of the ODMR spectrum.

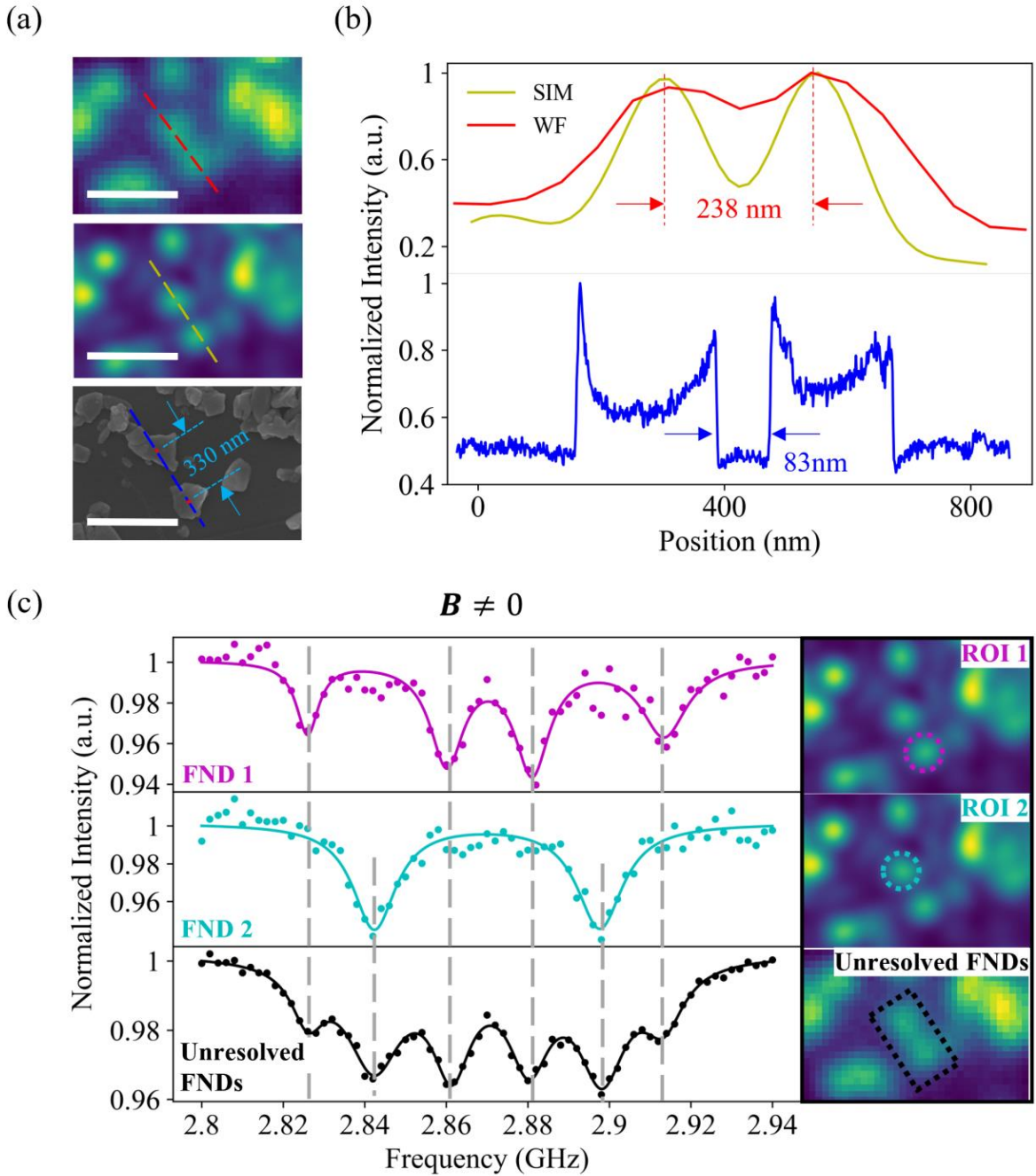

**Figure 5.** Super-resolved widefield ODMR measurements of the two unresolved FNDs. (a) WF, SIM and SEM images of the selected ROI. The scale bar in the inset is 500 nm. (b) Normal WF and SIM fluorescent intensity profiles of the FNDs along the red and yellow lines in Figure 5a,

respectively. The blue line is the SEM profile along the blue line in the SEM image in Figure 5a. (c) ODMR spectra extracted from the corresponding FNDs in the selected ROI (under an external magnetic field). The solid curves are the Lorentz fit of the ODMR data.

The ODMR performances under wide-field microscope and SIM have been analyzed in Table S1. The FWHM of the ODMR peaks under SIM was decreased compared with the FWHM under a wide-field microscope, while the contrast of the ODMR peaks under SIM was greater than the equivalent contrast under a wide-field microscope. The super-resolved ODMR measurements without external magnetic field exhibited the same trend of change in the linewidth and contrast as in the presence of an external magnetic field (Figure S2). This indicates that the super-resolution process could improve the sensitivity of measurements in magnetometry and thermometry, which are based on measuring the ODMR spectrums.

**Discussion**

It has been demonstrated that the long laser exposure may induce significant phototoxic side effects in living bio-samples from cells to organisms. [13, 14] This is particularly important for WQDM, as the laser power can reach the order of 10 kW/cm$^2$ [25, 26] which might create harsh conditions for the long-term illumination of cells. This issue could be significantly mitigated by using our dynamic illumination solution in WQDM, i.e., by "turning off" the unnecessary illumination in the region of the cell body (shown in Fig 2). By controlling the illumination, the overall exposure of the sample to excitation light can be effectively reduced.

NV-based widefield quantum sensing has been widely used for high sensitivity metrology with nanoscale spatial resolution, so our findings may result in important applications in a range of fields such as cellular biology, electronics and material science. [22, 47-49] It has been amply demonstrated [50] that the sensitivity of NV-based thermometry and magnetometry is highly correlated with the measured contrast and FWHM of the ODMR spectrum. This relationship will directly benefit from our new method, as our super-resolved ODMR measurements (Figure 5c) showed improved contrast and narrower linewidths than those obtainable using conventional methods. We also noticed the rapid development of SIM, and the adopted post-processing of data could be updated to real-time reconstruction using GPU-acceleration. [51, 52]

**Conclusion**

By using dynamic SLI, we have achieved a general super-resolved quantum sensing protocol. Specifically, the super-resolution process improves the performance of ODMR measurement, which leads to the sensitivity improvement of magnetic measurement. Through the dynamic SLI and variable illumination patterns, we can avoid phototoxicity and photobleaching caused by long-term laser exposure. The application of SLI in quantum sensing presents a new solution for precision ODMR measurements beyond the diffraction limit.

**Supporting Information**

Fluorescence nano-diamond sample preparation; Cell culture and fluorescent staining; Experimental setup; Super-resolved ODMR measurement sequence; Table S1 showing the FWHM and contrast of each ODMR peaks in Figure 5b; Figure S1 showing the measurement protocol for achieving super-resolved ODMR measurement; Figure S2 showing the super-resolved widefield ODMR measurements of FNDs without external magnetic field.

**Notes**

The authors declare no competing financial interest.

**Acknowledgments**

Z.Q.C. acknowledges financial support from the HKSAR Research Grants Council (RGC) Research Matching Grant Scheme (RMGS, No. 207300313) and HKSAR Innovation and Technology Fund (ITF) through the Platform Projects of the Innovation and Technology Support Program (ITSP; No. ITS/293/19FP). S.C.C acknowledges the support from the InnoHK Centre projects funded by the Innovation and Technology Commission (COCHE-1.5). The authors are grateful to Mr. Ye Tian (ME, HKU) for help in 3D printing of optical components.

**References**

(1) Degen, C. L.; Reinhard, F.; Cappellaro, P. Quantum sensing. *Reviews of Modern Physics* 2017, *89* (3), 035002.

(2) Doherty, M. W.; Manson, N. B.; Delaney, P.; Jelezko, F.; Wrachtrup, J.; Hollenberg, L. C. L. The nitrogen-vacancy colour centre in diamond. *Physics Reports* 2013, *528* (1), 1-45.

(3) Taylor, J. M.; Cappellaro, P.; Childress, L.; Jiang, L.; Budker, D.; Hemmer, P. R.; Yacoby, A.; Walsworth, R.; Lukin, M. D. High-sensitivity diamond magnetometer with nanoscale resolution. *Nature Physics* 2008, *4* (10), 810-816.

(4) Balasubramanian, G.; Chan, I. Y.; Kolesov, R.; Al-Hmoud, M.; Tisler, J.; Shin, C.; Kim, C.; Wojcik, A.; Hemmer, P. R.; Krueger, A.; et al. Nanoscale imaging magnetometry with diamond spins under ambient conditions. *Nature* 2008, *455* (7213), 648-651.

(5) Schirhagl, R.; Chang, K.; Loretz, M.; Degen, C. L. Nitrogen-Vacancy Centers in Diamond: Nanoscale Sensors for Physics and Biology. *Annual Review of Physical Chemistry* 2014, *65* (1), 83-105.

(6) Le Sage, D.; Arai, K.; Glenn, D. R.; DeVience, S. J.; Pham, L. M.; Rahn-Lee, L.; Lukin, M. D.; Yacoby, A.; Komeili, A.; Walsworth, R. L. Optical magnetic imaging of living cells. *Nature* 2013, *496* (7446), 486-489.

(7) Zhang, T.; Pramanik, G.; Zhang, K.; Gulka, M.; Wang, L.; Jing, J.; Xu, F.; Li, Z.; Wei, Q.; Cigler, P.; et al. Toward Quantitative Bio-sensing with Nitrogen–Vacancy Center in Diamond. *ACS Sensors* 2021, *6* (6), 2077-2107.

(8) Casola, F.; van der Sar, T.; Yacoby, A. Probing condensed matter physics with magnetometry based on nitrogen-vacancy centres in diamond. *Nature Reviews Materials* 2018, *3* (1), 17088.

(9) Steinert, S.; Ziem, F.; Hall, L. T.; Zappe, A.; Schweikert, M.; Götz, N.; Aird, A.; Balasubramanian, G.; Hollenberg, L.; Wrachtrup, J. Magnetic spin imaging under ambient conditions with sub-cellular resolution. *Nature Communications* 2013, *4* (1), 1607.

(10) Le Sage, D.; Arai, K.; Glenn, D. R.; DeVience, S. J.; Pham, L. M.; Rahn-Lee, L.; Lukin, M. D.; Yacoby, A.; Komeili, A.; Walsworth, R. L. Optical magnetic imaging of living cells. *Nature* 2013, *496* (7446), 486-489.

(11) Scholten, S. C.; Healey, A. J.; Robertson, I. O.; Abrahams, G. J.; Broadway, D. A.; Tetienne, J.-P. Widefield quantum microscopy with nitrogen-vacancy centers in diamond: Strengths, limitations, and prospects. *Journal of Applied Physics* 2021, *130* (15), 150902.

(12) Ascoli, G. A.; Bezhanskaya, J.; Tsytsarev, V. Microscopy. In *Encyclopedia of the Neurological Sciences (Second Edition)*, Aminoff, M. J., Daroff, R. B. Eds.; Academic Press, 2014; pp 16-20.

(13) Hoebe, R. A.; Van Oven, C. H.; Gadella, T. W. J.; Dhonukshe, P. B.; Van Noorden, C. J. F.; Manders, E. M. M. Controlled light-exposure microscopy reduces photobleaching and phototoxicity in fluorescence live-cell imaging. *Nature Biotechnology* 2007, *25* (2), 249-253.

(14) Laissue, P. P.; Alghamdi, R. A.; Tomancak, P.; Reynaud, E. G.; Shroff, H. Assessing phototoxicity in live fluorescence imaging. *Nature Methods* 2017, *14* (7), 657-661.

(15) Chakrova, N.; Canton, A. S.; Danelon, C.; Stallinga, S.; Rieger, B. Adaptive illumination reduces photobleaching in structured illumination microscopy. *Biomed. Opt. Express* 2016, *7* (10), 4263-4274.

(16) Pinkard, H.; Baghdassarian, H.; Mujal, A.; Roberts, E.; Hu, K. H.; Friedman, D. H.; Malenica, I.; Shagam, T.; Fries, A.; Corbin, K.; et al. Learned adaptive multiphoton illumination microscopy for large-scale immune response imaging. *Nature Communications* 2021, *12* (1), 1916.

(17) Hofmeister, A.; Thalhammer, G.; Ritsch-Marte, M.; Jesacher, A. Adaptive illumination for optimal image quality in phase contrast microscopy. *Optics Communications* 2020, *459*, 124972.

(18) Wilding, D.; Pozzi, P.; Soloviev, O.; Vdovin, G.; Verhaegen, M. Adaptive illumination based on direct wavefront sensing in a light-sheet fluorescence microscope. *Opt Express* 2016, *24* (22), 24896-24906.

(19) Ji, N. Adaptive optical fluorescence microscopy. *Nature Methods* 2017, *14* (4), 374-380.

(20) E., A. Beiträge zur Theorie des Mikroskops und der mikroskopischen Wahrnehmung. *Arkiv. Mikroskop. Anat.* 1873, *9*, 413-468.

(21) Dovzhenko, Y.; Casola, F.; Schlotter, S.; Zhou, T. X.; Büttner, F.; Walsworth, R. L.; Beach, G. S. D.; Yacoby, A. Magnetostatic twists in room-temperature skyrmions explored by nitrogen-vacancy center spin texture reconstruction. *Nature Communications* 2018, *9* (1), 2712.

(22) Kucsko, G.; Maurer, P. C.; Yao, N. Y.; Kubo, M.; Noh, H. J.; Lo, P. K.; Park, H.; Lukin, M. D. Nanometre-scale thermometry in a living cell. *Nature* 2013, *500* (7460), 54-58.

(23) Tam, J.; Merino, D. Stochastic optical reconstruction microscopy (STORM) in comparison with stimulated emission depletion (STED) and other imaging methods. *Journal of Neurochemistry* 2015, *135* (4), 643-658.

(24) Heintzmann, R.; Huser, T. Super-Resolution Structured Illumination Microscopy. *Chemical Reviews* 2017, *117* (23), 13890-13908.

(25) Pfender, M.; Aslam, N.; Waldherr, G.; Neumann, P.; Wrachtrup, J. Single-spin stochastic optical reconstruction microscopy. *Proceedings of the National Academy of Sciences* 2014, *111* (41), 14669-14674.

(26) Chen, E. H.; Gaathon, O.; Trusheim, M. E.; Englund, D. Wide-Field Multispectral Super-Resolution Imaging Using Spin-Dependent Fluorescence in Nanodiamonds. *Nano Letters* 2013, *13* (5), 2073-2077.

(27) Hirano, Y.; Matsuda, A.; Hiraoka, Y. Recent advancements in structured-illumination microscopy toward live-cell imaging. *Microscopy* 2015, *64* (4), 237-249.

(28) Wu, Y.; Shroff, H. Faster, sharper, and deeper: structured illumination microscopy for biological imaging. *Nature Methods* 2018, *15* (12), 1011-1019.

(29) Chen, J.; Fu, Z.; Chen, B.; Chen, S.-C. Fast 3D super-resolution imaging using a digital micromirror device and binary holography. *Journal of Biomedical Optics* 2021, *26* (11), 116502.

(30) Gustafsson, M. G. Surpassing the lateral resolution limit by a factor of two using structured illumination microscopy. *J Microsc* 2000, *198* (Pt 2), 82-87.

(31) Lal, A.; Shan, C.; Xi, P. Structured Illumination Microscopy Image Reconstruction Algorithm. *IEEE Journal of Selected Topics in Quantum Electronics* 2016, *22* (4), 50-63.

(32) Rainer, H.; Christoph, G. C. Laterally modulated excitation microscopy: improvement of resolution by using a diffraction grating. In *Proc.SPIE*, 1999; Vol. 3568, pp 185-196.

(33) Shao, L.; Kner, P.; Rego, E. H.; Gustafsson, M. G. L. Super-resolution 3D microscopy of live whole cells using structured illumination. *Nature Methods* 2011, *8* (12), 1044-1046.

(34) Meng, Y.; Lin, W.; Li, C.; Chen, S.-c. Fast two-snapshot structured illumination for temporal focusing microscopy with enhanced axial resolution. *Opt. Express* 2017, *25* (19), 23109-23121.

(35) Chen, J.; Gu, S.; Meng, Y.; Fu, Z.; Chen, S.-C. Holography-based structured light illumination for temporal focusing microscopy. *Opt. Lett.* 2021, *46* (13), 3143-3146.

(36) Ouyang, W.; Xu, X.; Lu, W.; Zhao, N.; Han, F.; Chen, S.-C. Ultrafast 3D nanofabrication via digital holography. *Nature Communications* 2023, *14* (1), 1716.

(37) Han, F.; Gu, S.; Klimas, A.; Zhao, N.; Zhao, Y.; Chen, S.-C. Three-dimensional nanofabrication via ultrafast laser patterning and kinetically regulated material assembly. *Science* 2022, *378* (6626), 1325-1331.

(38) Cheng, J.; Gu, C.; Zhang, D.; Chen, S.-C. High-speed femtosecond laser beam shaping based on binary holography using a digital micromirror device. *Opt. Lett.* 2015, *40* (21), 4875-4878.

(39) Cheng, J.; Gu, C.; Zhang, D.; Wang, D.; Chen, S.-C. Ultrafast axial scanning for two-photon microscopy via a digital micromirror device and binary holography. *Opt. Lett.* 2016, *41* (7), 1451-1454.

(40) Fu, Z.; Geng, Q.; Chen, J.; Chu, L.-A.; Chiang, A.-S.; Chen, S.-C. Light field microscopy based on structured light illumination. *Opt. Lett.* 2021, *46* (14), 3424-3427.

(41) Fu, Z.; Chen, J.; Liu, G.; Chen, S.-C. Single-shot optical sectioning microscopy based on structured illumination. *Opt. Lett.* 2022, *47* (4), 814-817.

(42) Geng, Q.; Gu, C.; Cheng, J.; Chen, S.-c. Digital micromirror device-based two-photon microscopy for three-dimensional and random-access imaging. *Optica* 2017, *4* (6), 674-677.

(43) Wen, C.; Ren, M.; Feng, F.; Chen, W.; Chen, S.-C. Compressive sensing for fast 3-D and random-access two-photon microscopy. *Opt. Lett.* 2019, *44* (17), 4343-4346.

(44) Chen, D.; Ren, M.; Zhang, D.; Chen, J.; Gu, S.; Chen, S.-C. Design of a multi-modality DMD-based two-photon microscope system. *Opt. Express* 2020, *28* (20), 30187-30198.

(45) Chen, D.; Chen, B.; Shao, Q.; Chen, S.-C. Broadband angular dispersion compensation for digital micromirror devices. *Opt. Lett.* 2022, *47* (3), 457-460.

(46) Chu, Z.; Zhang, S.; Zhang, B.; Zhang, C.; Fang, C.-Y.; Rehor, I.; Cigler, P.; Chang, H.-C.; Lin, G.; Liu, R.; et al. Unambiguous observation of shape effects on cellular fate of nanoparticles. *Scientific Reports* 2014, *4* (1), 4495.

(47) Dolde, F.; Fedder, H.; Doherty, M. W.; Nöbauer, T.; Rempp, F.; Balasubramanian, G.; Wolf, T.; Reinhard, F.; Hollenberg, L. C. L.; Jelezko, F.; et al. Electric-field sensing using single diamond spins. *Nature Physics* 2011, *7* (6), 459-463.

(48) Dadras, S.; Mukherjee, A.; Shayan, K.; Tarduno, J. A.; Vamivakas, A. N. Wide-Field Magnetic and Thermal Imaging of Electric Currents Using NV− Centers in Nanodiamond Ensembles. In *2021 Conference on Lasers and Electro-Optics (CLEO)*, 9-14 May 2021, 2021; pp 1-2.

(49) Foy, C.; Zhang, L.; Trusheim, M. E.; Bagnall, K. R.; Walsh, M.; Wang, E. N.; Englund, D. R. Wide-Field Magnetic Field and Temperature Imaging Using Nanoscale Quantum Sensors. *ACS Applied Materials & Interfaces* 2020, *12* (23), 26525-26533.

(50) Dréau, A.; Lesik, M.; Rondin, L.; Spinicelli, P.; Arcizet, O.; Roch, J. F.; Jacques, V. Avoiding power broadening in optically detected magnetic resonance of single NV defects for enhanced dc magnetic field sensitivity. *Physical Review B* 2011, *84* (19), 195204.

(51) Jin, L.; Liu, B.; Zhao, F.; Hahn, S.; Dong, B.; Song, R.; Elston, T. C.; Xu, Y.; Hahn, K. M. Deep learning enables structured illumination microscopy with low light levels and enhanced speed. *Nature Communications* 2020, *11* (1), 1934.

(52) Markwirth, A.; Lachetta, M.; Mönkemöller, V.; Heintzmann, R.; Hübner, W.; Huser, T.; Müller, M. Video-rate multi-color structured illumination microscopy with simultaneous real-time reconstruction. *Nature Communications* 2019, *10* (1), 4315.